\documentclass[twocolumn,english,aps,prl,superscriptaddress,reprint,showpacs]{revtex4}
\usepackage{mathptmx}

\usepackage[T1]{fontenc}
\usepackage[latin9]{inputenc}
\usepackage{amsmath}
\usepackage{amssymb}
\usepackage{graphicx}
\usepackage{esint}

\makeatletter
\@ifundefined{textcolor}{}
{%
 \definecolor{BLACK}{gray}{0}
 \definecolor{WHITE}{gray}{1}
 \definecolor{RED}{rgb}{1,0,0}
 \definecolor{GREEN}{rgb}{0,1,0}
 \definecolor{BLUE}{rgb}{0,0,1}
 \definecolor{CYAN}{cmyk}{1,0,0,0}
 \definecolor{MAGENTA}{cmyk}{0,1,0,0}
 \definecolor{YELLOW}{cmyk}{0,0,1,0}
 }

\@ifundefined{definecolor}{\usepackage{color}}{}

\DeclareMathOperator{\e}{e}

\usepackage{babel}

\usepackage{babel}

\usepackage{babel}

\usepackage{babel}

\usepackage{babel}

\makeatother

\usepackage{babel}
\begin{document}

\title{Universal Superfluid Transition and Transport Properties of Two-Dimensional
Dirty Bosons}

\author{Giuseppe Carleo}

\affiliation{Laboratoire Charles Fabry, Institut d'Optique, CNRS, Univ Paris Sud
11, 2 avenue Augustin Fresnel, F-91127 Palaiseau cedex, France}

\author{Guilhem Boéris}

\affiliation{Laboratoire Charles Fabry, Institut d'Optique, CNRS, Univ Paris Sud
11, 2 avenue Augustin Fresnel, F-91127 Palaiseau cedex, France}

\author{Markus Holzmann}

\affiliation{LPTMC, UMR 7600 of CNRS, Université Pierre et Marie Curie, 75005
Paris, France}

\affiliation{Université Grenoble 1/CNRS, LPMMC, UMR 5493, B.P. 166, 38042 Grenoble,
France}

\author{Laurent Sanchez-Palencia}

\affiliation{Laboratoire Charles Fabry, Institut d'Optique, CNRS, Univ Paris Sud
11, 2 avenue Augustin Fresnel, F-91127 Palaiseau cedex, France}
\begin{abstract}
We study the phase diagram of two-dimensional, interacting bosons
in the presence of a correlated disorder in continuous space, using
large-scale finite temperature quantum Monte Carlo simulations. We
show that the superfluid transition is strongly protected against
disorder. It remains of the Berezinskii-Kosterlitz-Thouless type up
to disorder strengths comparable to the chemical potential. Moreover,
we study the transport properties in the strong disorder regime where
a zero-temperature Bose-glass phase is expected. We show that the
conductance exhibits a thermally activated behavior vanishing only
at zero temperature. Our results point towards the existence of Bose
\emph{bad-metal} phase as a precursor of the Bose-glass phase. 
\end{abstract}


\maketitle

\paragraph*{Introduction~--}

Transport properties in quantum materials are governed by a complex
interplay of disorder and interactions. While disorder tends to localize
particles, interactions may strongly alter the single-particle picture
by either reinforcing or suppressing localization. Dramatic effects
are expected in two dimensions (2D), such as metal-insulator transitions~\cite{abrahams2001,tracy2009,allison2006},
suppression of superfluidity~\cite{crowell1995,crowell1997,luhman2004,luhman2006,lazarowich2006},
and presumably high-$T_{\textrm{c}}$ superconductivity~\cite{pan2001}.
Possible phase transitions are however particularly elusive owing
to absence of true long-range order even in extended phases~\cite{mermin1966,hohenberg1967},
and many questions remain open. For instance, resistance measurements
in Si-MOFSETs suggest a metal-insulator transition~\cite{abrahams2001},
which may be attributed to quantum localization or classical percolation~\cite{tracy2009},
but experiments on GaAs heterostructures point towards a crossover
behavior~\cite{allison2006}. Studies of the superfluid Berezinskii-Kosterlitz-Thouless
(BKT) transition~\cite{Berez1971,0022-3719-6-7-010} have also been
reported for $^{4}$He films adsorbed on porous media~\cite{crowell1995,crowell1997}.
While the BKT transition is unaffected in the weak disorder limit~\cite{0022-3719-7-9-009},
the question of its relevance for strong disorder is left open owing
to the difficulty to identify a universal jump of the superfluid density~\cite{luhman2004,luhman2006,lazarowich2006}.
Moreover, a question that is attracting much debate is whether many-body
localization effects~\cite{Basko2006,oganesyan2007,PhysRevLett.99.180402,PhysRevA.84.013612,Aleiner2010}
can drive a finite-temperature metal-insulator transition in two dimensions.

Ultracold quantum gases in controlled disorder offer a unique tool
to address these questions in a unified way~\cite{LSPNature2010}.
In clean or disordered quasi-2D geometries, direct consequences of
vortex pairing~\cite{Hadzibabic2006}, superfluidity~\cite{desbuquois2012},
quasi-long-range phase coherence~\cite{Beeler2012,PhysRevA.85.033602},
and resistance measurements~\cite{krinner2013} have been reported.
On the theoretical side, most knowledge rely on lattice models with
uncorrelated disorder~\cite{alet2003,prokofev2004,wysin2005,soyler2011,lin2011,Zuniga2013}.
Conversely, little is known when the disorder is continuous, correlated
and can sustain a classical percolation threshold. An important class
of such models is the one realized by optical speckle potentials used
in recent ultracold-atom experiments ~\cite{1367-2630-8-8-165,pezze2011b,PhysRevLett.102.150402,Pilati2010,PhysRevA.86.063626}.

In this Letter, we report on an \textit{ab-initio} path-integral quantum
Monte Carlo study of the phase diagram of interacting 2D bosons in
the presence of a speckle-type disorder. We show that, although the
density profile exhibits large spatial modulations, the superfluid
BKT transition is strongly protected against disorder. This holds
up to disorder strengths comparable to the chemical potential, where
the zero-temperature Bose-glass transition is expected. The critical
properties of the \emph{dirty} superfluid transition can be understood
in terms of a universal description including a simple renormalization
of the critical parameters. In particular, we find that the superfluid
transition occurs while the fluid percolates with a density significantly
above the critical density of the clean system. It allows us to rule
out the classical percolation scenario. Moreover, we study the conductance
by means of a novel quantum Monte Carlo estimator. Deep in the strong
disorder regime, we find direct evidence of an insulating behavior,
characterized by a conductance that decreases with temperature. Our
data is consistent with a thermally-activated behavior of the Arrhenius
type, indicating that the conductance vanishes only at zero temperature.

\paragraph*{System and methods~--}

We consider a two-dimensional quantum fluid of interacting bosons
of mass $m$ at temperature $T$ subjected to a disordered potential
$V(\mathbf{r})$, governed by the Hamiltonian 
\begin{equation}
H=\sum_{i}\left[-\frac{\nabla_{i}^{2}}{2m}+V(\mathbf{r}_{i})\right]+\sum_{i<j}u(\mathbf{r}_{i}-\mathbf{r}_{j}).
\end{equation}
 The properties of the short-range repulsive interaction potential
$u(r)$ used in ultracold-atom experiments can be accurately described
by a s-wave pseudopotential with an effective coupling strength $g$
\cite{PhysRevLett.84.2551}%
\footnote{The effective two-dimensional coupling parameter in ultra-cold atoms
experiment reads $g=\sqrt{8\pi}\frac{a_{s}}{a_{z}}\frac{\hbar^{2}}{m}$,
where $a_{s}$ is the three-dimensional s-wave scattering length and
$a_{z}=\sqrt{\hbar/m\omega_{z}}$ is the characteristic trapping length
in the transverse direction. %
}. The disordered potential we take is a correlated, isotropic, and
continuous speckle potential, as realized by laser light diffusion
through a ground-glass plate~\cite{1367-2630-8-8-165,Beeler2012,PhysRevA.85.033602,krinner2013}.
Its average value coincides with its standard deviation $V_{\text{\tiny R}}$,
and its two-point correlation function is a Gaussian of r.m.s.\ radius
$\sigma_{\text{\tiny R}}$ \cite{goodman2007speckle}%
\footnote{The disordered potential is obtained as the square modulus of the
discrete Fourier transform of $\frac{\sqrt{2\pi V_{\text{\tiny R}}}}{\tilde{\sigma}L}\exp\left[-\frac{k_{x}{}^{2}+k_{y}{}^{2}}{4\tilde{\sigma}^{2}}+i\theta(k_{x},k_{y})\right]$,
where $\tilde{\sigma}^{2}=\frac{1}{2\sigma_{\text{\tiny R}}^{2}}$,
$\theta(k_{x},k_{y})$ is an uniformly distributed random phase and
$L$ is the linear size of the two-dimensional system. The wave-vectors
are integer multiples of $\Delta k=\frac{2\pi}{L}$ and we typically
take $\sim10^{4}$ wave-vectors along each spatial direction. %
}. In the following, we consider an interaction strength $\tilde{g}\equiv mg/\hbar^{2}=0.1$,
a fixed chemical potential $\mu$ and a correlation radius $\sigma_{\text{\tiny R}}=\sqrt{5}\xi_{0}$,
where $\xi_{0}=\hbar/\sqrt{2m\mu}$ is the healing length in the absence
of disorder. These are typical values realized in ultracold-atom experiments~\cite{Beeler2012,PhysRevA.85.033602}.
For instance, for $^{87}\text{Rb}$ atoms the chosen chemical potential
corresponds to $\mu/k_{\text{\tiny B}}\simeq56\,$nK and $\sigma_{\text{\tiny R}}=0.5\,\mu$m.

Our study is based on a fully \emph{ab-initio} path-integral quantum
Monte Carlo (QMC) approach, which allows for unbiased calculations
of equilibrium, finite-temperature properties of interacting bosons~\cite{RevModPhys.67.279}
in terms of a discretized path-integral representation using $M$
time slices. We use the worm algorithm~\cite{PhysRevLett.96.070601}
to sample the grand-canonical partition function at inverse temperature
$\beta=1/k_{\tiny\textnormal{B}}T$. Whereas disorder is taken into
account at the level of the primitive-approximation, the two-body
interaction is treated in the the pair-product approximation~\cite{RevModPhys.67.279,PhysRevLett.77.3695}.
The two-particle propagator is derived by solving the s-wave scattering
problem at high temperatures, $\epsilon^{-1}=Mk_{\tiny\textnormal{B}}T$,
which yields 
\begin{multline}
\langle\mathbf{r}|\e^{-\epsilon H_{2}}|\mathbf{r^{\prime}}\rangle\simeq\frac{m}{4\pi\hbar^{2}\epsilon}\e^{-m\left|\mathbf{r}-\mathbf{r^{\prime}}\right|^{2}/4\hbar^{2}\epsilon}\\
+\frac{\tilde{g}}{8\pi}\int_{0}^{\infty}dk\, ke^{-\epsilon\frac{\hbar^{2}k^{2}}{m}}\left[J_{0}(kr)Y_{0}(kr')+J_{0}(kr')Y_{0}(kr)\right],\label{eq:twopartprop}
\end{multline}
 where $H_{2}$ is the Hamiltonian of the relative motion of two particles,
$\mathbf{r}$ and $\mathbf{r}^{\prime}$ are the relative coordinates
at different times, whereas $J_{0}$ and $Y_{0}$ are Bessel functions
of the first and second kind, respectively. Contributions of higher
($l>0$) partial waves to the two-particle density matrix are approximated
by their non-interacting expressions. Here, we report results from
simulations of up to $N=10^{5}$ particles in a box of linear extension
$L$ with periodic boundary conditions, averaged over $\approx40$
disorder realizations. The typical number of time slices used here
vary from $M=16$ for weak disorder to $M=70$ for strong disorder
amplitudes.

\paragraph*{Superfluid transition~--}

We first study the superfluid to normal fluid transition in the presence
of disorder. To elucidate the critical properties of this transition,
we calculate the static superfluid susceptibility, $\chi_{\text{s}}=\frac{1}{L^{2}}\int d\mathbf{r}d\mathbf{r}^{\prime}\left\langle \Psi^{\dagger}(\mathbf{r})\Psi(\mathbf{r}^{\prime})\right\rangle $,
and the field correlation function, $g_{1}(r)=\left\langle \Psi^{\dagger}(\mathbf{r})\Psi(0)\right\rangle $,
where $\Psi$ is the field operator, and $\left\langle \dots\right\rangle $
denotes thermal and disorder average. In the superfluid phase of the
clean system the decay of field correlations is algebraic, $g_{1}(r)\sim r^{-\eta}$,
and the superfluid susceptibility diverges as $\chi_{\text{s}}\sim L^{2-\eta}$.
The exponent $\eta$ is directly related to the superfluid density
$n_{\text{s}}$ as $\eta=\frac{mk_{\text{\tiny B}}T}{2\pi\hbar^{2}n_{\text{s}}}$.
In the normal phase, field correlations are exponentially suppressed,
$g_{1}(r)\sim\e^{-r/\gamma}$, and the susceptibility remains finite
in the thermodynamic limit.

\begin{figure}
\noindent \begin{centering}
\includegraphics[width=1\columnwidth]{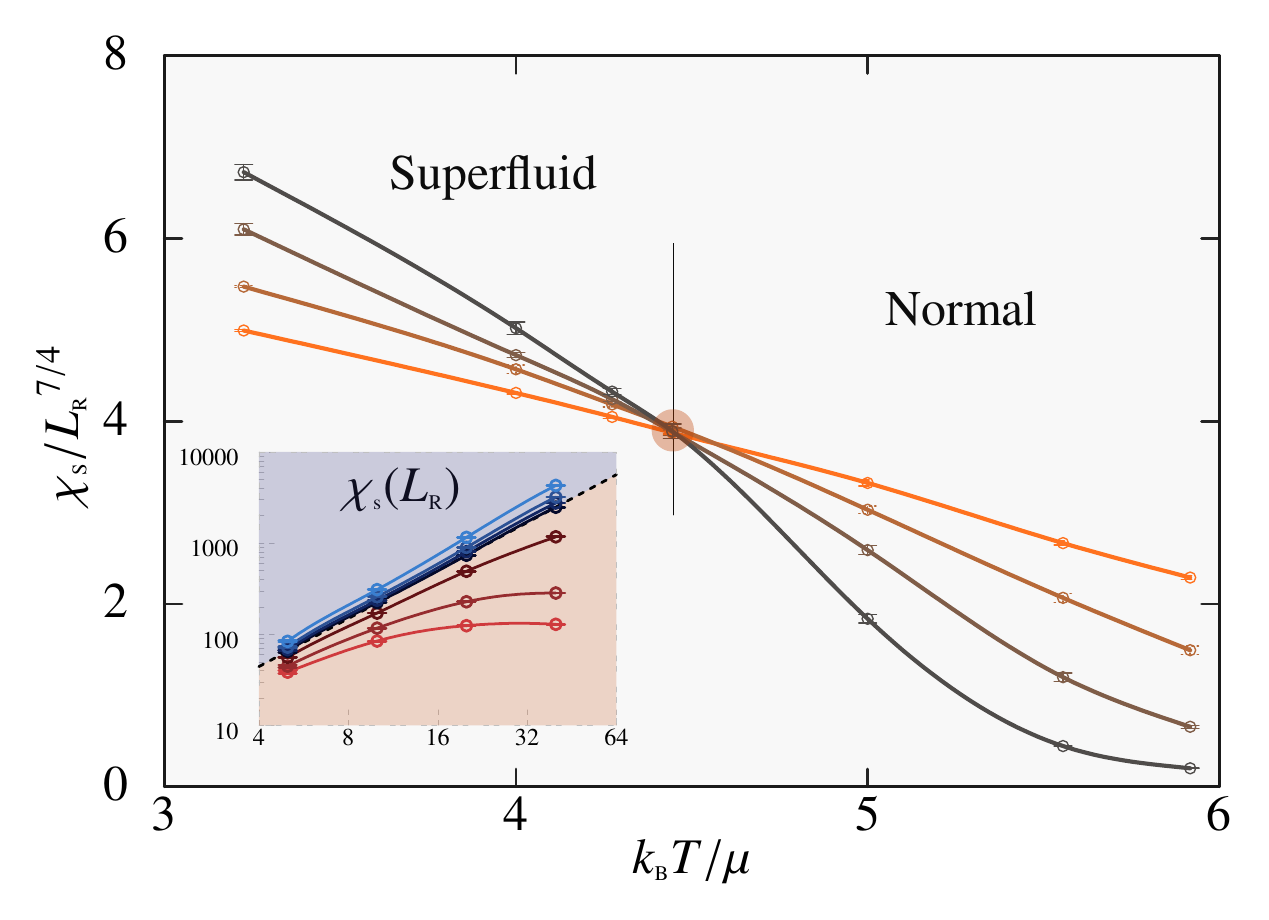} 
\par\end{centering}

\caption{Scaled superfluid susceptibility against temperature for system sizes
ranging from $L_{\tiny\textnormal{R}}=L/\sigma_{\tiny\textnormal{R}}=5-40$
at disorder strength $V_{\tiny\text{R}}/\mu=0.3$ and interaction
strength $\tilde{g}=0.1$. Darker curves mark increasingly large size.
In the inset, the susceptibility is shown as a function of $L_{\tiny\textnormal{R}}$
(temperature increases from top to bottom). The divergence predicted
by the BKT transition for clean systems ($\eta=1/4$) is indicated
by the dashed line. \label{fig:Scaling-of-the}}
\end{figure}

Analysis of our QMC data allows us to study both qualitatively and
quantitatively the superfluid to normal transition in the presence
of disorder (Fig.~\ref{fig:Scaling-of-the}). The inset of Fig. \ref{fig:Scaling-of-the}
shows the superfluid susceptibility, $\chi_{\text{s}}$, as a function
of the system size for various temperatures at intermediate disorder
amplitude, $V_{\textnormal{\tiny R}}/\mu=0.3$. For high temperatures
(curves in the light-orange zone), $\chi_{\text{s}}$ converges to
a finite value, characteristic for the normal fluid phase. Above the
critical temperature (curves in the blue zone), $\chi_{\text{s}}$
shows an algebraic divergence. At the critical point, our QMC results
are compatible with the scaling $\chi_{\text{s}}\sim L^{7/4}$, i.e.
$\eta\simeq1/4$, as expected for the BKT transition in the clean
system \cite{Berez1971,0022-3719-6-7-010}. It is confirmed by the
behavior of the rescaled superfluid susceptibility $\chi_{\text{s}}/L^{7/4}$,
plotted in the main panel of Fig.~\ref{fig:Scaling-of-the} as a
function of temperature for various values of $L$, where all curves
cross nearly at a single point. Further, our data for the superfluid
density calculated from the winding number estimator \cite{RevModPhys.67.279}
(not shown) is consistent with the universal jump at the transition
temperature in the thermodynamic limit, expected for a BKT transition.
We conclude that the superfluid transition of dirty bosons remains
in the universality class of the clean BKT transition. To some extent,
this result is expected for weak disorder. According to Harris argument
\cite{0022-3719-7-9-009}, local fluctuations of the disorder potential
are smoothed out at the scale of the diverging correlation length
at the BKT transition, thus introducing only a renormalization of
the effective parameters. Our calculations show that it holds also
for strong disorder (up to $V_{\text{\tiny R}}\simeq\mu$).

The intersection point of the rescaled superfluid susceptibility (curves
as in Fig.~\ref{fig:Scaling-of-the}) precisely determines the transition
temperature at fixed disorder amplitude. In our analysis we find that
the BKT scaling regime is attained for \emph{increasingly large} system
sizes, upon increasing the disorder strength. It results in a residual
size dependence of the susceptibility intersection points. The latter
is nonetheless well behaved and amenable for $1/L$ extrapolation
of the transition temperature in the thermodynamic limit. In order
to have sufficiently small finite-size corrections to the the BKT
scaling, we use, for the largest disorder strength reported, systems
of linear size $L/\sigma_{\tiny\text{R}}\sim40-80$. These sizes should
be contrasted with the typically (much) smaller ones of the inner
critical regions probed in experiments \cite{luhman2004,luhman2006,lazarowich2006,PhysRevA.85.033602,Beeler2012}
and in some numerical simulations \cite{Astra2013}.

\begin{figure}
\noindent \begin{centering}
\includegraphics[width=1\columnwidth]{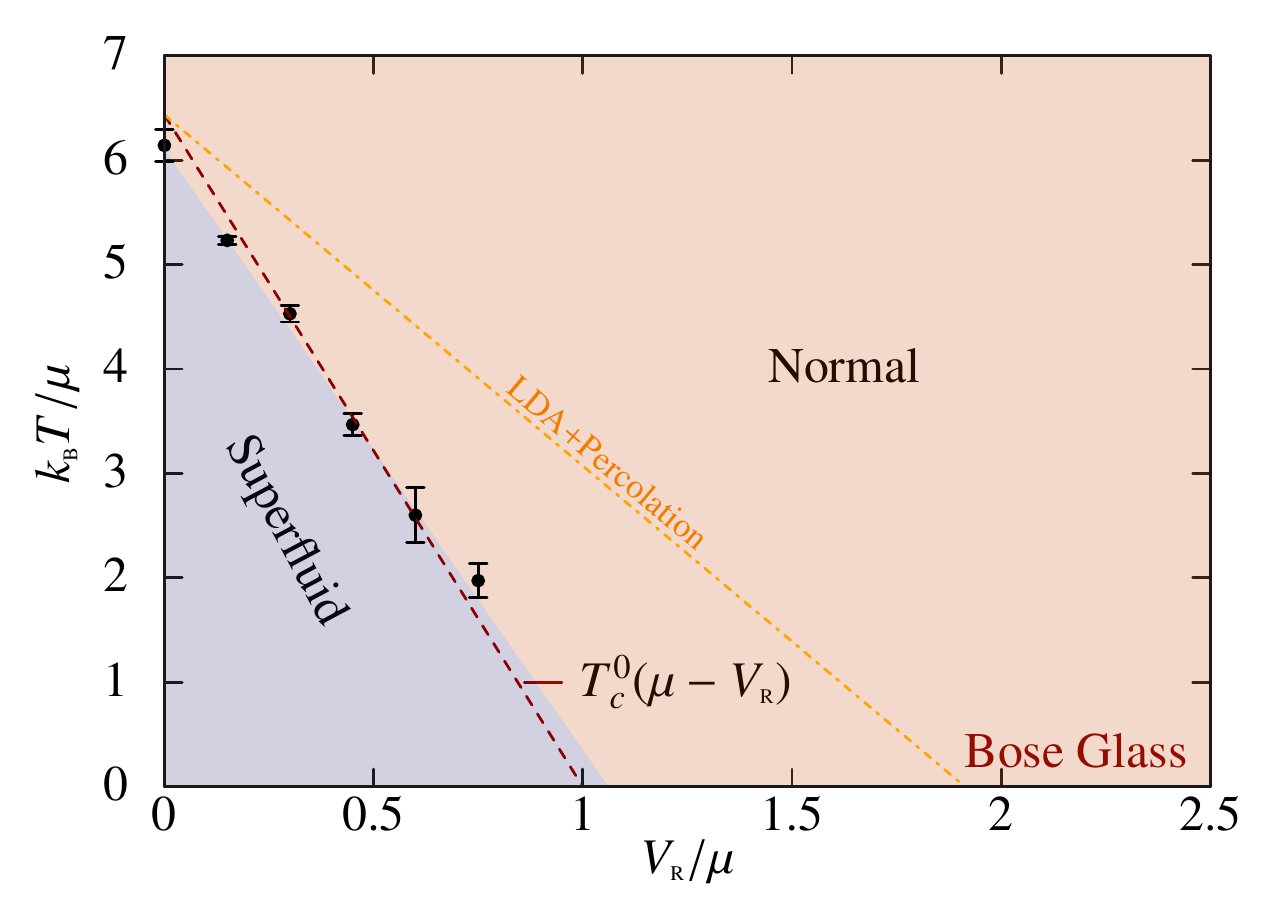} 
\par\end{centering}

\caption{Phase diagram of two-dimensional interacting bosons at fixed chemical
potential and interaction strength $\tilde{g}=0.1$. The dark dashed
line is the critical temperature of the clean system with a renormalized
chemical potential; the lighter dashed line is the prediction of a
combination of LDA and percolation theory (see text). For strong disorder,
the normal system goes to the Bose-glass phase in the zero-$T$ limit.\label{fig:Phase-diagram-of}}
\end{figure}

In Fig.~\ref{fig:Phase-diagram-of}, we show the resulting critical
temperatures versus disorder strength and draw the superfluid to normal
fluid phase diagram of dirty 2D bosons. Within numerical accuracy,
the critical line is described by $T_{\text{c}}(V_{\tiny\text{R}})\simeq T_{\text{c}}^{0}(\mu-V_{\tiny\text{R}})$,
where $T_{c}^{0}\simeq\frac{\pi\mu}{\tilde{g}k_{\text{B}}}/\log(13.2/\tilde{g})$
is the critical temperature for the disorder-free system~\cite{PhysRevLett.87.270402}.
The shift of the critical temperature can be understood in terms of
the leading order renormalization of the chemical potential, $\mu_{c}(V_{\tiny\text{R}})-V_{\tiny\text{R}}=\mu_{c}(0)$.
For our parameters, upon increasing the disorder strength the critical
density decreases from the disorder-free value $n_{c}\sigma_{\tiny\text{R}}^{2}\simeq21$
to the zero-temperature ($V_{\tiny\text{R}}\simeq\mu$) linearly-extrapolated
value of $n_{c}\sigma_{\tiny\text{R}}^{2}\simeq7$. At lower densities,
the system remains normal for any of our temperatures, $k_{\textnormal{B}}T\gtrsim\mu$,
consistent with a Bose-glass phase at zero temperature, as indicated
in Fig.~\ref{fig:Phase-diagram-of}.

We now compare our results to the possibility of a percolation-driven
transition, as occurring in classical systems and investigated both
experimentally \cite{PhysRevB.20.3653} and theoretically \cite{pezze2011b}
for 2D speckle disorder. In the superfluid phase, the Bose gas density
for each realization of the disorder may be approximated using the
local density approximation (LDA), $n_{\textnormal{\tiny LDA}}(\mathbf{r},V_{\text{R}})=n_{0}\left[\mu-V(\mathbf{r})\right]$,
where $n_{0}(\mu)$ is the density of the clean system at chemical
potential $\mu$. The quantitative agreement between this approximation
and our QMC results is reasonably good in the superfluid phase. We
consider percolating clusters in which the local density stays everywhere
higher than the critical density of the clean system, i.e. $n_{\textnormal{\tiny LDA}}(\mathbf{r},V_{\text{R}})\gtrsim\frac{mk_{\text{\tiny B}}T}{2\pi\hbar^{2}}\log\left(380/\tilde{g}\right)$
\cite{PhysRevLett.87.270402}. The occurrence of such percolating
superfluid clusters yields the orange line reported in Fig. \ref{fig:Phase-diagram-of},
which shows that the percolation scenario strongly overestimates the
critical disorder strength. Superfluidity disappears while there still
exist large percolating clusters above the critical density of the
clean system. Notice that we have disregarded the weak penetration
of the density into the disorder potential barriers due to the finite
healing length in the superfluid phase~\cite{PhysRevA.74.053625}.
However, this effect does not alter the above conclusions since it
would further raise the percolation line in Fig.~\ref{fig:Phase-diagram-of}.

\paragraph*{Transport Properties --}

We now address the strongly disordered regime, where superfluid coherence
is lost and disorder substantially affects mass transport. A long-debated
issue is to understand whether interacting systems preserve Anderson-like
localization properties, as found in the single particle case \cite{Basko2006,oganesyan2007,PhysRevLett.99.180402,PhysRevA.84.013612,Aleiner2010}.
At zero temperature, the conditions for a localized (Bose glass) phase
to exist have been established in seminal works \cite{PhysRevB.37.325,PhysRevB.40.546}.
At finite temperature, however, the existence of a perfectly insulating
state, characterized by an exactly vanishing conductance ($G_{\text{\tiny DC}}=0$)
is instrinsically more complex and theoretically plausible scenarios
have been put forward only recently. The \emph{many-body localization}
transition scenario \cite{Basko2006,Aleiner2010} implies the existence
of a critical temperature, $T_{\text{loc}}$, separating the perfect
insulator behavior at low energies where $G_{\text{\tiny DC}}(T<T_{\text{loc}})=0$,
from a delocalized, diffusive phase with $G_{\text{\tiny DC}}(T>T_{\text{loc}})\neq0$.
This transition, however, has not been directly observed in experiments
and remains at the center of intense theoretical activity, for instance
in out-of-equilibrium phenomena \cite{PhysRevB.83.094431,Carleo2012,1367-2630-14-9-095020}.

In order to identify possible signatures of many-body localization
at finite temperature, we have computed transport properties in the
strongly disordered region. Such transport properties are traditionally
addressed in disordered materials subjected to an external electric
bias. They were recently shown to be also accessible in a new generation
of ultracold-atom experiments \cite{krinner2013}. 
\begin{figure}[b]
\noindent \begin{centering}
\includegraphics[width=1\columnwidth]{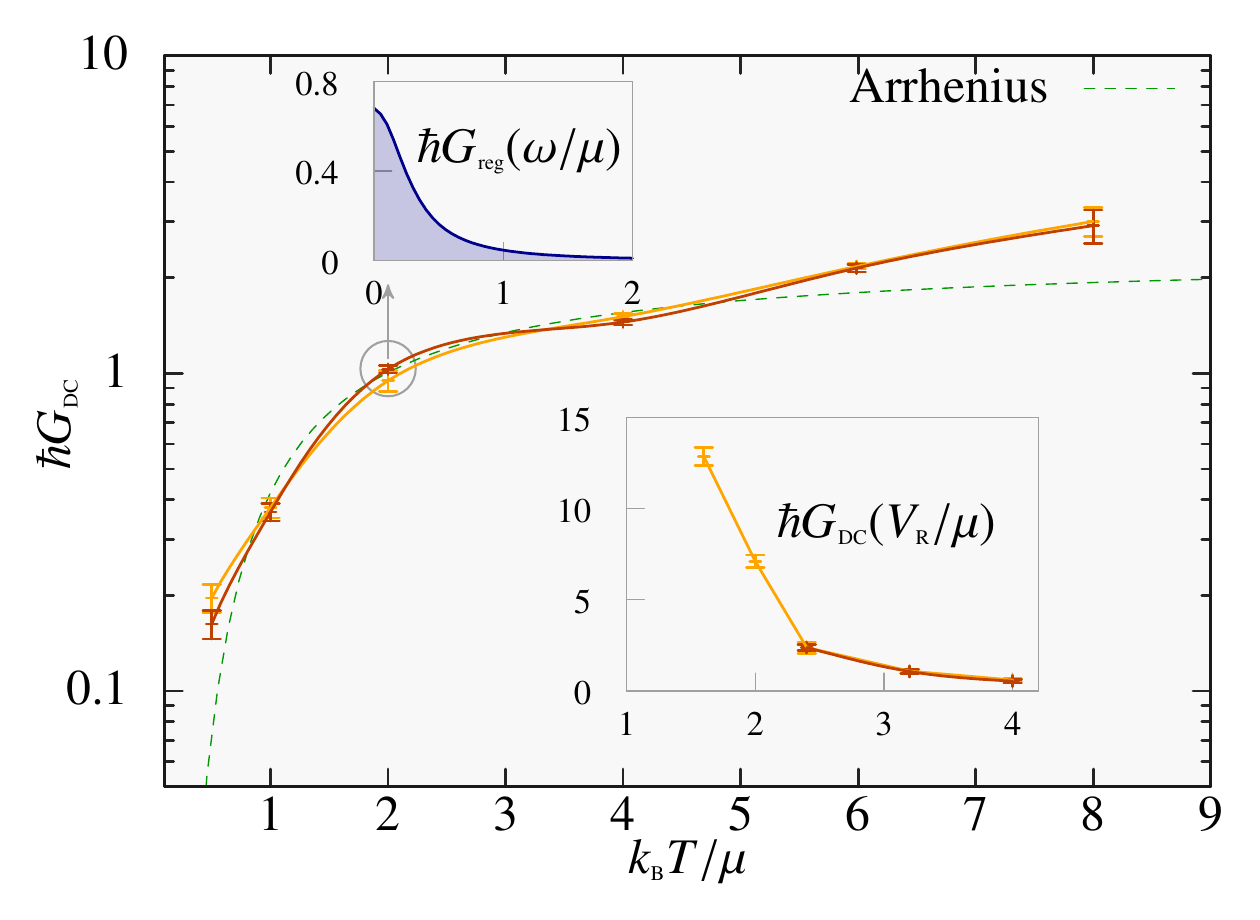} 
\par\end{centering}

\caption{Finite-temperature, zero-frequency longitudinal conductance for an
intensity of the disorder $V_{\tiny\text{R}}=2.4\ \mu$. Error bars
refer to averages over disorder and the dashed line is a fit to a
simple exponential Arrhenius law $G_{\text{\tiny DC}}(T)\propto\left(\frac{\mu}{k_{\textnormal{\tiny B}}T}\right)e^{-\frac{\Delta E}{k_{\text{\tiny B}}T}}$.
In the upper inset, the complete frequency dependence of the conductance
is exemplified for $k_{\textrm{\tiny B}}T=2\mu$. In the lower inset,
the conductance at constant temperature ($k_{\textnormal{\tiny B}}T=\mu$)
and varying disorder strength is also shown. In all cases darker curves
mark increasingly large sizes, in the range $L/\sigma_{\tiny\textnormal{R}}=20-40$.
\label{fig:Finite-temperature-conductivity}}
\end{figure}

To obtain the zero-frequency (DC) conductance within QMC, we compute
the imaginary-time correlations of the density current operator along
a given direction $\alpha$, $\Lambda_{\alpha}(\tau,q)=\langle J_{\alpha}(\tau,q)J_{\alpha}(0,-q)\rangle$.
It allows us to reconstruct the associated longitudinal conductance
$G_{\text{reg}}(\omega)$ %
\footnote{Notice that we refer here to mass transport. The conventional electric
conductance is $\tilde{G}_{\tiny\textnormal{reg}}(\omega)=q^{2}G_{\textnormal{reg}}(\omega)$
for particles of charge $q$. %
} from numerical analytical continuation \cite{PhysRevB.57.10287,PhysRevB.44.6011,PhysRevB.80.094301}
of the Laplace transform \cite{Kubo1966} 
\begin{eqnarray}
\Lambda_{\alpha}(\tau,0) & = & 2\hbar{\displaystyle \int_{-\infty}^{+\infty}d\omega\frac{\exp\left(-\omega\tau\right)\omega}{1-\exp\left(-\beta\hbar\omega\right)}G_{\text{reg}}(\omega).}\label{eq:fluctdiss}
\end{eqnarray}
 Imaginary-time correlations of high statistical quality are essential
for a reliable reconstruction of the dynamical response $G_{\text{reg}}(\omega)$.
Direct estimators of current correlations successfully used for lattice
simulations \cite{PhysRevB.47.7995,PhysRevB.54.R3756} present a diverging
statistical error in the small imaginary-time step limit in continuum
systems, preventing its use for numerical analytical continuation
for our purposes. To overcome this serious problem, we introduce a
novel estimator that significantly reduces the statistical error making
possible a reliable determination of the conductance. The key observation
is that the imaginary-time derivatives of the correlations are well-behaved
in the small time-step limit $\epsilon\rightarrow0$, and that the
imaginary-time integral of the correlation function is also well-behaved
and directly related to the superfluid density $n_{\textnormal{s}}$
\cite{PhysRevB.36.8343}. We can therefore express the current-current
correlations by means of the integral expression 
\begin{multline}
\Lambda_{\alpha}(\tau,0)=\frac{1}{\beta m}\left(n-n_{\textnormal{s}}\right)+\int_{0}^{\beta}d\tau_{1}\Lambda_{\alpha}^{\prime}(\tau_{1},0)+\\
-\frac{1}{\beta}\int_{0}^{\beta}d\tau_{1}\int_{0}^{\tau_{1}}d\tau_{2}\Lambda_{\alpha}^{\prime}(\tau_{2},0),\label{eq:lambdafin}
\end{multline}
 where the imaginary-time derivative of the correlation function reads
\begin{multline}
\epsilon\Lambda_{\alpha}^{\prime}(\tau,0)=\frac{1}{mL^{2}}\sum_{i}\left\langle \left[\nabla_{i}^{\alpha}V(\tau)\right]\times D^{\alpha}(0)\right\rangle +\\
-\frac{1}{2mL^{2}}\sum_{i,j,\beta}\left\langle D_{j}^{\beta}(\tau)\times\left[\nabla_{i}^{\alpha}\nabla_{j}^{\beta}V(\tau)\right]\times D^{\alpha}(0)\right\rangle +\mathcal{O}(\epsilon^{\frac{3}{2}}).\label{eq:lambdader}
\end{multline}
 In the latter expression $\nabla_{i}^{\alpha}$ is the derivative
with respect to the position of the $i$-th particle, $D^{\alpha}(\tau)=\sum_{j}^{N}D_{j}^{\alpha}(\tau)=\sum_{j}^{N}\left[R_{j}^{\alpha}(\tau)-R_{j}^{\alpha}(\tau-\epsilon)\right]$
is proportional to the imaginary-time displacement of the center of
mass and $V(\tau)$ is a generic one-body potential which in our specific
case is the disordered speckle potential.

We first study the behavior of the DC longitudinal conductance $G_{\text{DC}}=G_{\text{reg}}(\omega=0)$
versus the temperature deep in the strong disorder regime. The outcome
is shown in the main panel of Fig. \ref{fig:Finite-temperature-conductivity}
for a fixed disorder strength, $V_{\tiny\textnormal{R}}=2.4\,\mu$.
The conductance monotonously decreases with temperature, a behavior
which is typically observed in related experiments with disordered
materials. In the zero-temperature limit, the longitudinal conductance
is consistent with a vanishing value, as expected for the Bose-glass
phase \cite{PhysRevB.40.546}. The overall behavior of $G_{\text{\tiny DC}}$,
however, points towards a thermally activated transport (see fit in
Fig. \ref{fig:Finite-temperature-conductivity}), at variance with
a perfectly insulating, many-body localized phase at finite temperature,
at least on the temperature scale accessible in our study. The energy
broadening of the reconstructed dynamical conductances (see upper
inset of Fig. \ref{fig:Finite-temperature-conductivity}) remains
finite and does not show any indication for a finite-temperature mobility
edge.

In the lower inset of Fig. \ref{fig:Finite-temperature-conductivity}
we further show the behavior of the longitudinal conductance at fixed
temperature versus the disorder strength $V_{\textnormal{\tiny R}}$,
ranging from the edge of the superfluid transition $V_{\textnormal{\tiny R}}\sim\mu$
to the strong disorder regime, with $V_{\textnormal{\tiny R}}\sim4\mu$.
Compatibly with the thermally activated scenario, we find that the
transport is exponentially suppressed with increasing disorder, but
never vanishes, in agreement with experimental results \cite{krinner2013}.
Size effects, though present, do not indicate any phase transition,
at least for mesoscopic samples. Therefore, our \emph{ab-initio }analysis
hints to a delocalized \emph{bad Bose-metal} phase at finite-temperature.

\paragraph*{Conclusions -- }

We have studied two-dimensional interacting bosons in the presence
of a correlated disordered potential using \emph{ab-initio} quantum
Monte Carlo calculations. We have found that the disorder renormalizes
the low-energy Hamiltonian in the neighborhood of the superfluid transition,
but the critical line remains in the universality class of the Berezinkii-Kosterlitz-Thouless
transition. It holds for arbitrary strong disorder up the zero-temperature
Bose-glass transition. Moreover, we have developed a new estimator
for the conductance, which does not suffer from diverging statistical
errors. Deep in the strong disorder regime, the mass transport exhibits
a thermally activated behavior and it is strictly suppressed only
at zero temperature. It points towards the existence of a \emph{Bose
bad-metal} phase, as the finite-temperature precursor of the Bose-glass
insulator. We have found no indication of a finite-temperature many-body
localization transition. Our results provide new theoretical insights
both in the general understanding of disordered interacting systems
and in the interpretation of experiments with both ultracold atoms
and disordered materials.

\paragraph*{Acknowledgments --}

We acknowledge discussions with B.~Altshuler, D.~Basko, T.~Bourdel,
T.~Giamarchi, T.~Roscilde, S.~Moroni, and T. Ziman. This research
was supported by the European Research Council (FP7/2007-2013 Grant
Agreement No.~256294), Ministère de l'Enseignement Supérieur et de
la Recherche, and Agence Nationale de la Recherche ({}``DisorderTransitions'').
It was performed using HPC resources from GENCI-CCRT/CINES (Grants
t2012056853 and t2013056853). Use of the computing facility cluster
GMPCS of the LUMAT federation (FR LUMAT 2764) is also aknowledged.
Our simulations make use of the ALPS~%
\footnote{http://alps.comp-phys.org/%
} scheduler library and statistical analysis tools~\cite{Troyer1998,Albuquerque20071187,1742-5468-2011-05-P05001}.

\end{document}